\documentclass[10pt, conference, letterpaper]{IEEEtran}
\IEEEoverridecommandlockouts
\usepackage{dsfont}
\usepackage{cite}
\usepackage{amsmath,amssymb,amsfonts}
\usepackage{algorithmic}
\usepackage{graphicx}
\graphicspath{{figures/}}
\usepackage{textcomp}
\usepackage{wasysym} 
\usepackage{adjustbox}
\usepackage{xcolor}
\usepackage{enumerate} 
\usepackage{booktabs}
\usepackage{enumitem}
\usepackage{multirow}
\usepackage{xspace}
\usepackage{array}
\usepackage{tabularx}
\usepackage{subcaption}
\usepackage{colortbl}
\usepackage{pifont}
\usepackage{pgfplots}
\usepackage{tikz}
\usetikzlibrary{shapes, arrows.meta, positioning, calc, patterns, decorations.pathreplacing, fit, backgrounds}
\pgfplotsset{compat=1.18}
\usepackage{balance}
\usepackage[hidelinks]{hyperref}

\newcommand{\eg}{\textit{e.g.}}

\newcommand{\ourmethod}{\texttt{WhiteNet}\xspace}

\def\BibTeX{{\rm B\kern-.05em{\sc i\kern-.025em b}\kern-.08em
    T\kern-.1667em\lower.7ex\hbox{E}\kern-.125emX}}

\tikzset{
    block/.style={draw, fill=blue!8, rounded corners=2pt, minimum height=0.55cm, minimum width=1.2cm, font=\scriptsize, align=center, line width=0.4pt},
    blockG/.style={draw, fill=green!10, rounded corners=2pt, minimum height=0.55cm, minimum width=1.2cm, font=\scriptsize, align=center, line width=0.4pt},
    blockR/.style={draw, fill=red!8, rounded corners=2pt, minimum height=0.55cm, minimum width=1.2cm, font=\scriptsize, align=center, line width=0.4pt},
    blockO/.style={draw, fill=orange!12, rounded corners=2pt, minimum height=0.55cm, minimum width=1.2cm, font=\scriptsize, align=center, line width=0.4pt},
    blockP/.style={draw, fill=violet!10, rounded corners=2pt, minimum height=0.55cm, minimum width=1.2cm, font=\scriptsize, align=center, line width=0.4pt},
    arrow/.style={-{Stealth[length=1.5mm]}, line width=0.35pt},
    darrow/.style={<->{Stealth[length=1.5mm]}, line width=0.35pt},
    skipline/.style={dashed, gray, line width=0.3pt},
}

\makeatletter
\def\ps@IEEEtitlepagestyle{%
  \def\@oddfoot{\mbox{}\hfill
    \raisebox{2mm}{\fbox{\parbox{0.97\textwidth}{
    \small
      This work has been submitted to the IEEE for possible publication. Copyright may be transferred without notice, after which this version may no longer be accessible.}}}%
    \hfill\mbox{}}%
  \def\@evenfoot{\@oddfoot}%
}
\makeatother

\begin{document}

\title{\ourmethod:~Robust Identification of Overlapping IEEE 802.11 Signals Across Unseen Channels}


\author{
  \IEEEauthorblockN{Ildi Alla\textsuperscript{*}, Vincent Lenders\textsuperscript{*}}
  \IEEEauthorblockA{\textsuperscript{*}SnT, University of Luxembourg, Luxembourg \\
  \{ildi.alla, vincent.lenders\}@uni.lu}
}

\maketitle

\begin{abstract}
Deep learning (DL) classifiers trained on I/Q samples achieve high accuracy for IEEE 802.11 protocol identification of overlapping signals, but their performance degrades sharply when channel conditions at deployment differ from those encountered during training.
We present \ourmethod{}, a framework that addresses the problem of  \emph{channel variability} in I/Q samples. The central idea is \emph{spectral whitening}, a physics-grounded preprocessing step that suppresses frequency-selective fading while preserving protocol-discriminative features. 
To reduce dependence on costly multi-transmitter over-the-air captures for training, we complement it with a \emph{synthetic overlap mixer} featuring a physically accurate per-transmitter channel and shared-receiver signal chain for pre-training without extensive field data collection.
On public over-the-air IEEE 802.11 data, \ourmethod{} closes a substantial portion of the accuracy gap caused by unseen channel conditions while using 7.7 times fewer parameters than the prior state of the art, and optionally distills to compact variants for coarse spectrum awareness on power-constrained edge devices.
\end{abstract}


\vspace{-0.35cm}
\section{Introduction}
\vspace{-0.2cm}
\label{sec:introduction}

The coexistence of multiple IEEE~802.11 protocol generations in shared ISM spectrum, from the direct-sequence spread spectrum (DSSS) of 802.11b to the orthogonal frequency-division multiple access (OFDMA) of 802.11ax, creates an increasingly complex electromagnetic environment~\cite{9864321}. When two or more Wi-Fi transmissions occupy the same frequency band simultaneously, their signals superimpose into composite waveforms that conventional preamble-based detectors cannot decompose~\cite{belgiovine2024tprime, zhang2021signal}. Real-time identification of the \emph{constituent protocol versions} within such \emph{spectrally overlapping} signals is a prerequisite for cognitive spectrum access~\cite{khalek2023advances}, interference management~\cite{szott2022wi, 8325299}, and the detection of unauthorized transmissions in critical infrastructure~\cite{zhang2025physical, alla2024robust}.

\begin{figure}[t]
    \centering
    \captionsetup{font=small}
    \includegraphics[width=\columnwidth]
    {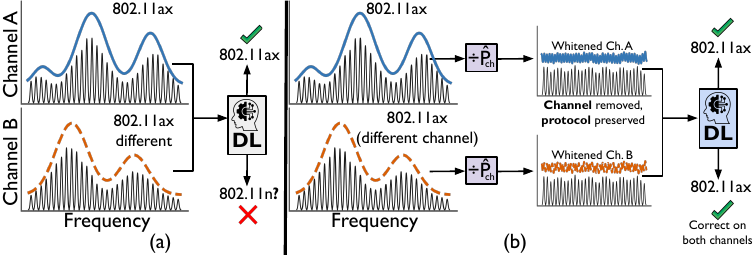}
    \vspace{-0.55cm}
    \caption{The channel robustness gap. (a)~Different channel conditions reshape the spectrum of the same protocol, causing misclassification. (b)~Spectral whitening removes the channel envelope, preserving protocol-discriminative features.}
    \label{fig:motivation}
    \vspace{-0.8cm}
\end{figure}

Deep learning (DL) has emerged as a powerful tool for protocol-level identification directly from raw in-phase/quadrature (I/Q) samples \cite{jagannath2022comprehensive, 8737459}, bypassing the fragility of correlation-based preamble matching. Recent work spanning transformers~\cite{belgiovine2024tprime}, residual networks~\cite{oshea2018over}, attention-enhanced convolutional classifiers~\cite{zhang2022high, zhang2021attention_amc}, and joint detection-classification systems~\cite{9493767, xie2024joint_detection_amc} has demonstrated high accuracy on tasks ranging from single-protocol classification to overlapping Wi-Fi signals~\cite{belgiovine2024tprime}, as well as wideband spectrum segmentation~\cite{alla2026finding, uvaydov2024stitching}. However, these results rest on a critical, largely unexamined assumption: \emph{the channel conditions at deployment will match those encountered during training}. Models are trained and tested under the same multipath propagation conditions, noise floor, and receiver hardware, a setting that flatters accuracy but conceals a fundamental fragility (Fig.~\ref{fig:motivation}a).

In practice, each physical setting imprints a unique frequency-selective fading signature~$H(f)$ on every received signal, shaped by geometry, materials, and transceiver placement~\cite{rappaport2024wireless}. A classifier trained under one set of channel conditions learns decision boundaries entangled with those conditions. When conditions change, the spectral shapes shift unpredictably, causing severe performance degradation. Standard domain adaptation~\cite{ganin2016dann} offers limited relief, requiring many diverse training domains. With limited over-the-air (OTA) capture diversity, gradient reversal lacks sufficient signal. Meta-learning~\cite{nichol2018reptile} similarly needs numerous tasks, while few-shot approaches~\cite{mackey2022cross} require support data from target conditions. Prior work in radio frequency (RF) fingerprinting has documented similar channel-induced failures~\cite{al2020exposing, shen2023toward}, yet no existing method addresses channel-robust \emph{protocol classification under spectral overlap}. This gap has direct deployment consequences. A spectrum monitoring system whose channel conditions evolve over time would require periodic signal re-collection and model re-training to maintain reliable operation, limiting the scalability of \textit{data-driven spectrum intelligence}.


\newcommand{\halfcirc}{\scalebox{0.78}{\LEFTcircle}} 

\begin{table*}[t]
\centering
\captionsetup{font=small}
\caption{Positioning of \ourmethod{} relative to existing approaches ($\bullet$: supported, $\circ$: not supported, \halfcirc: partial, (--): not reported).}
\vspace{-0.2cm}
\label{tab:related_work}
\adjustbox{max width=\linewidth}{%
\begin{tabular}{@{}l l c c c c c c c@{}}
\toprule
\textbf{Work} & \textbf{Architecture} & \textbf{Parameters} & \textbf{Task} & \textbf{Overlap} & \textbf{Held-Out} & \textbf{Synthetic Overlap} & \textbf{Channel} & \textbf{Edge} \\
 &  & (M) &   & \textbf{Classification} & \textbf{Evaluation} & \textbf{Generation} & \textbf{Invariance} & \textbf{Deploy} \\
\midrule
\cite{oshea2018over, zhang2022high, li2024complex} & ResNet, HoCANs, CV-TRN & 0.229, 0.213, 0.171 & AMC & $\circ$ & $\circ$ & $\circ$ & $\circ$ & $\circ$ \\
\cite{bu2020adversarial}  & CNN+DA & -- & AMC & $\circ$ & \halfcirc\rlap{$^\dagger$} & $\circ$ & $\bullet$ & $\circ$ \\
\cite{mackey2022cross}  & PTN & -- & RFF & $\circ$ & $\bullet$\rlap{$^\ddagger$} & $\circ$ & $\bullet$ & $\circ$ \\
\cite{shen2023toward}  & Transformer & 0.349 & RFF & $\circ$ & $\circ$ & $\circ$ & $\bullet$ & $\circ$ \\
\cite{kong2024deepcrf}  & CNN+CL & 0.125 & RFF & $\circ$ & $\bullet$ & $\circ$ & $\bullet$ & $\circ$ \\
\cite{xie2024joint_detection_amc}  & CNN + YOLO & -- & Det. + AMC & \halfcirc\rlap{$^\S$} & $\circ$ & \halfcirc\rlap{$^\S$} & $\circ$ & $\circ$ \\
\cite{belgiovine2024tprime}  & Transformer & 6.8 & Protocol ID & $\bullet$ & \halfcirc\rlap{$^\|$} & $\circ$ & $\circ$ & $\bullet$ \\ 
\midrule
\textbf{\ourmethod{} (Ours)} & \textbf{U-Net+NL} & \textbf{0.889} & \textbf{Protocol ID} & $\bullet$ & $\bullet$ & $\bullet$ & $\bullet$ & $\bullet$ \\
\bottomrule
\multicolumn{9}{@{}l}{\rule{0pt}{2.2ex}\scriptsize 
$^\dagger$Domain shift via downsampling on the same dataset; not evaluated on held-out OTA captures.
$^\ddagger$Cross-domain evaluation in~\cite{mackey2022cross} requires few-shot support from target conditions.}\\
\multicolumn{9}{@{}l}{\scriptsize $^\S$Coexisting signals at distinct carrier frequencies; no same-band spectral overlap. $^\|$Single-protocol only; overlap not evaluated on unseen channels.
}
\end{tabular}}
\vspace{-0.7cm}
\end{table*}

Our key observation is that the physics of wireless propagation offers a direct path to channel invariance. Since IEEE 802.11 deployments are overwhelmingly indoor, the channel coherence bandwidth~$B_c$ will typically exceed 1~MHz~\cite{rappaport2024wireless}, meaning $|H(f)|$ varies \emph{slowly} across frequency. The modulation features that distinguish OFDM-based wireless protocols (subcarrier spacing, guard intervals, pilot patterns, resource unit boundaries) vary on the scale of tens to hundreds of kHz~\cite{ieee80211_2020}. This \emph{scale separation} enables a signal-processing intervention: \textbf{spectral whitening} divides each observation's spectrum by its smoothed power spectral density (PSD), flattening the slow-varying channel while preserving fast-varying protocol signatures (Fig.~\ref{fig:motivation}b). Building on this insight, we present \ourmethod{}, a compact framework comprising spectral whitening preprocessing, a physically grounded \emph{synthetic overlap mixer} that generates training data from single-protocol captures, and a U-Net encoder--decoder with non-local attention that compresses via knowledge distillation to as few as 10K parameters for edge deployment.

\vspace{0.07cm}
\noindent \textbf{Summary of Contributions}
\vspace{0.07cm}


\noindent $\bullet$ We introduce \emph{spectral whitening} as a preprocessing step for wireless protocol classification, exploiting the scale separation between channel coherence bandwidth and modulation subcarrier spacing. This physics-grounded intervention removes channel-specific fading without target-condition data, recovering +26.0\,pp exact-match accuracy on channels unseen during training and narrowing the in-distribution to held-out gap from 
40.0 to 8.0\,pp.


\noindent $\bullet$ We design a transmitter--channel--receiver simulation pipeline that generates realistic overlapping signals from single-protocol OTA captures, including per-transmitter multipath, carrier frequency offset (CFO), shared-oscillator phase noise, and receiver nonlinearity. Synthetic pre-training enables multi-label overlap classification, reducing dependence on costly multi-transmitter data collection and lifting held-out exact match from 47.4\% to 73.6\% over naive signal addition.


\vspace{-0.1cm}
\section{Related Work}
\vspace{-0.1cm}
\label{sec:related_work}

Table~\ref{tab:related_work} compares \ourmethod{} with existing approaches across five capabilities 
required by a deployable spectrum monitoring system: classification of overlapping 
co-channel signals, evaluation under channel conditions unseen during training (held-out), synthetic generation of overlapping signals for training, 
an explicit method for channel-invariant classification,
and compression for 
edge deployment.

\vspace{-0.13cm}
\subsection{Protocol Classification and Overlap Detection}
\vspace{-0.1cm}
DL has become the dominant paradigm for classifying wireless signals from raw I/Q samples~\cite{8357902}, with architectures ranging from residual networks~\cite{oshea2018over} and attention-enhanced classifiers~\cite{zhang2022high, zhang2021attention_amc} to complex-valued models~\cite{li2024complex}. These works focus on automatic modulation classification (AMC), distinguishing modulation families such as QAM or BPSK, and typically evaluate on synthetic channels at controlled  signal-to-noise ratio (SNR). The more challenging task of \emph{protocol-level} identification (\eg, 802.11ax vs.\ 802.11n sharing the same OFDM modulation) has received less attention. Recent transformer-based approaches~\cite{belgiovine2024tprime} achieve high performance on \emph{overlapping} Wi-Fi signals, and joint detection--classification frameworks~\cite{xie2024joint_detection_amc} classify multiple modulated signals coexisting at distinct carrier frequencies, while semantic spectrum segmentation~\cite{uvaydov2024stitching} localizes multiple coexisting technologies at per-frequency-bin granularity and multi-instance multi-label learning~\cite{pan2020automatic} handles overlapping radar waveforms. However, all these methods assume matched channel conditions between training and deployment, leaving robustness to unseen deployment channels largely unexplored.

\vspace{-0.1cm}
\subsection{Channel Robustness in RF Systems}
\vspace{-0.1cm}
The vulnerability of learned RF features to channel-induced distribution shift is well documented. The authors in~\cite{al2020exposing} showed that channel characteristics dominate learned representations, degrading RF fingerprinting (RFF) when channel conditions change~\cite{11556387}. Subsequent work has proposed length-versatile representations~\cite{shen2023toward}, prototypical networks for few-shot adaptation~\cite{mackey2022cross}, federated contrastive learning~\cite{shen2024federated}, and adversarial transfer~\cite{bu2020adversarial} to improve cross-domain robustness. More recently, contrastive learning with model-inspired channel augmentation~\cite{kong2024deepcrf} has demonstrated channel-resilient Wi-Fi device fingerprinting across physically distinct environments. These approaches have advanced the RF fingerprinting field considerably but require either labeled target-domain samples, many diverse training domains, few-shot support sets, device-level CSI features unavailable in wideband spectrum monitoring, or protocol-specific preamble structure with quasi-static narrowband channels~\cite{shen2023toward, 9450821}. None addresses the distinct challenge of \emph{overlapping protocol identification} under channel variation without target-condition data.

\begin{figure*}[t]
    \centering
    \vspace{-0.2cm}
    \captionsetup{font=small}
    \includegraphics[width=\linewidth]{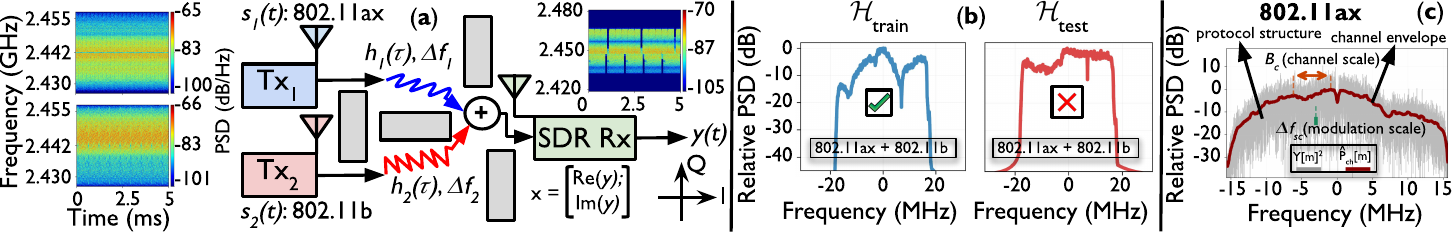}
    \caption{System model and problem illustration. (a)~$K$ transmitters through independent channels $h_k(\tau)$ to a wideband software-defined radio (SDR) receiver. (b)~Real OTA PSD of the same ax+b overlap pair under training (left) and held-out (right) channels. (c)~Scale separation: the smoothed envelope (red) tracks $|H(f)|^2$ while modulation structure (gray) is preserved at scale $\Delta\!f_{\text{sc}} \ll B_c$.}
    \label{fig:system_model}
    \vspace{-0.65cm}
\end{figure*}

\vspace{-0.1cm}
\subsection{Spectral Normalization}
\vspace{-0.1cm}
Spectral whitening
(dividing a signal's spectrum by its smoothed envelope) 
is a classical technique in adjacent fields: seismology~\cite{bensen2007processing} uses it to equalize the peaked ambient-noise spectrum and suppress persistent narrowband sources, while speech processing employs RASTA filtering~\cite{hermansky1994rasta} and cepstral mean normalization~\cite{atal1974effectiveness} to suppress channel effects.
In wireless communications, frequency-domain equalization is standard for demodulation~\cite{rappaport2024wireless}, but it requires knowledge of the transmitted signal (via pilots or preambles) and is applied \emph{after} detection, not as a preprocessing step for classification. Under co-channel overlap, equalization faces an additional barrier. The interfering signal corrupts pilot-based channel estimates~\cite{ieee80211_2020, perahia2013next}, and the protocol identity needed to locate the correct preamble fields is precisely the unknown being classified (Section~\ref{subsec:results}).
To the best of our knowledge, \textbf{\emph{no prior work has applied spectral whitening as a preprocessing step for wireless protocol classification}}.

\vspace{-0.1cm}
\section{System Model and Problem Formulation}
\label{sec:system_model}
\vspace{-0.12cm}
\subsection{Received Signal Model}
\vspace{-0.1cm}

We consider a wideband 
SDR receiver capturing a wide bandwidth in the ISM band, where up to $K$ transmitters may be simultaneously active, with or without spectral overlap (Fig.~\ref{fig:system_model}a). Each transmitter~$k$ emits a baseband signal $s_k(t)$ conforming to one of $|\mathcal{P}|$ wireless protocol classes $\mathcal{P}$. Without loss of generality, we instantiate $\mathcal{P}$ with four IEEE~802.11 generations, $\mathcal{P} = \{\text{ax}, \text{b}, \text{n}, \text{g}\}$, whose physical layers span single-carrier DSSS, OFDM, and OFDMA-based transmission. The received complex baseband signal is:
\vspace{-0.25cm}
\begin{equation}
y(t) = \sum_{k=1}^{K} \alpha_k \big(h_k * s_k\big)(t)\, e^{j2\pi \Delta\!f_k t} + v(t),
\label{eq:signal_model}
\vspace{-0.15cm}
\end{equation}
where $h_k(\tau)$ is the multipath channel impulse response between transmitter~$k$ and the receiver,
$\alpha_k \in \mathbb{R}^{+}$ is the received amplitude scaling, 
$\Delta\!f_k$ is the CFO due to oscillator mismatch at transmitter~$k$, and $v(t)$ is complex AWGN whose samples satisfy $v[n] \sim \mathcal{CN}(0, \sigma_v^2)$ after digitization. 
The channel frequency response $H_k(f)$ exhibits \emph{frequency-selective fading} whose fluctuation rate is governed by the \emph{coherence bandwidth} $B_c \approx 1/(5\sigma_\tau)$, where $\sigma_\tau$ is the RMS delay spread~\cite{rappaport2024wireless}. 
In the frequency domain, the channel response takes the form:
\vspace{-0.25cm}
\begin{equation}
H_k(f) = \sum_{\ell=0}^{L-1} a_{k,\ell}\, e^{-j2\pi f \tau_{k,\ell}},
\label{eq:channel_freq}
\vspace{-0.25cm}
\end{equation}
where $a_{k,\ell} \in \mathbb{C}$ and $\tau_{k,\ell}$ are the complex gain and excess 
delay of the $\ell$-th multipath component, and $L$ is the number of resolvable paths.
Crucially, $H_k(f)$ is determined by the propagation environment 
(geometry, materials, and transceiver positions), so different channel conditions produce different $H_k(f)$ for identical transmitted signals.

\begin{figure*}[t]
  \centering
  \captionsetup{font=small}
  \includegraphics[width=\linewidth]{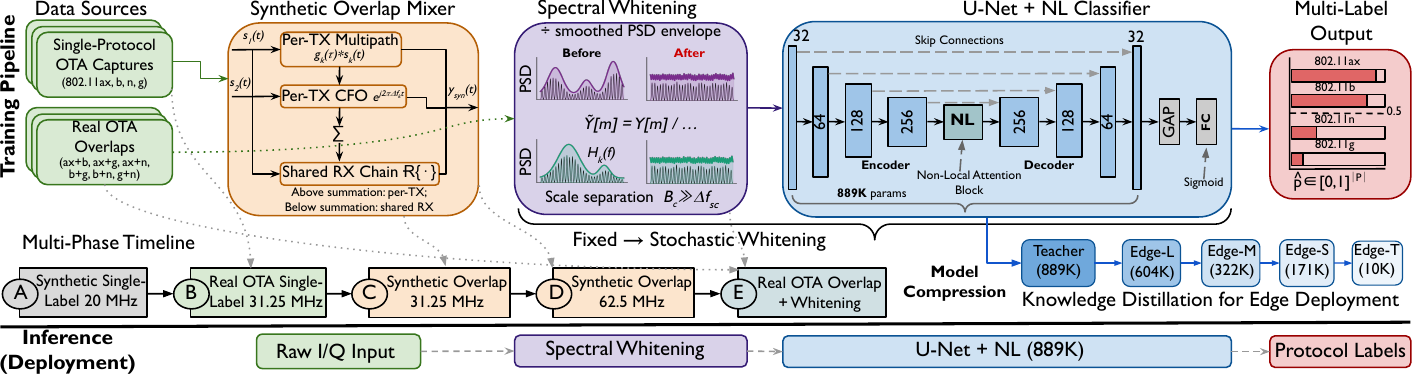}
  \vspace{-0.55cm}
  \caption{Overview of 
  \ourmethod{}. 
  Single-protocol OTA captures are synthetically mixed with per-transmitter channel impairments, then spectrally whitened before multi-label classification. At inference, only fixed whitening and a single forward pass are required.}
  \vspace{-0.65cm}
  \label{fig:framework}
\end{figure*}

\vspace{-0.15cm}
\subsection{Classification and the Channel Robustness Problem}
\vspace{-0.1cm}
\label{subsec:cross_env}

In the frequency domain, the received signal spectrum at frequency $f$ can be expressed (using $H_k(f{-}\Delta\!f_k) \approx H_k(f)$, valid since $|\Delta\!f_k| \ll B_c$) as:
\vspace{-0.25cm}
\begin{equation}
Y(f) = \sum_{k=1}^{K} \alpha_k \, H_k(f) \, S_k(f{-}\Delta\!f_k) + V(f),
\label{eq:freq_domain}
\vspace{-0.2cm}
\end{equation}
where $S_k(f)$ carries protocol-specific spectral signatures: DSSS spreading for 802.11b, OFDM subcarrier patterns for 802.11g/n, and OFDMA resource unit structure for 802.11ax~\cite{ieee80211_2020}. 
The classification task is formulated as a \emph{multi-label} problem. Given the digitized observation $\mathbf{y} = [y[0], y[1], \ldots, y[N{-}1]]^T \in \mathbb{C}^N$, represented as a two-channel real input $\mathbf{x} = [\text{Re}(\mathbf{y});\, \text{Im}(\mathbf{y})] \in \mathbb{R}^{2 \times N}$, a classifier $\mathcal{M}_\theta: \mathbb{R}^{2 \times N} \to [0,1]^{|\mathcal{P}|}$ parameterized by $\theta$ predicts the probability of each protocol being present:
\vspace{-0.35cm}
\begin{equation}
\hat{\mathbf{p}} = \mathcal{M}_\theta(\mathbf{x}), \quad \hat{p}_i = \sigma\!\big(z_i(\mathbf{x};\theta)\big), \quad i \in \{1, \ldots, |\mathcal{P}|\},
\label{eq:classifier}
\vspace{-0.3cm}
\end{equation}
where $z_i$ is the $i$-th output logit and $\sigma(\cdot)$ denotes the sigmoid function. 
The model is trained to minimize the binary cross-entropy loss $\mathcal{L}_{\text{BCE}} = -\sum_{i} \big[l_i \log \hat{p}_i + (1{-}l_i)\log(1{-}\hat{p}_i)\big]$, where $l_i \in \{0,1\}$ is the ground-truth presence indicator.

The fundamental challenge is that $\mathcal{M}_\theta$ never observes $S_k(f)$ directly. It observes $S_k(f)$ \emph{modulated by} the channel $H_k(f)$. Let $\mathcal{H}_{\text{train}}$ denote the distribution of channel realizations seen during training. A model trained under $\mathcal{H}_{\text{train}}$ learns decision boundaries shaped by those channel statistics. When deployed under a different channel distribution $\mathcal{H}_{\text{test}} \neq \mathcal{H}_{\text{train}}$, whether from temporal variation in the same setting or from an entirely different propagation environment, the spectral shapes shift unpredictably due to the new channel responses, causing misclassification, as shown in Fig.~\ref{fig:system_model}b.

\vspace{-0.2cm}
\subsection{Scale Separation: A Path to Channel Invariance}
\vspace{-0.1cm}
\label{subsec:scale_separation}

The central observation motivating our approach is that $H_k(f)$ and the protocol-discriminative structure of $S_k(f)$ operate at fundamentally different frequency scales. The channel varies slowly, with coherence bandwidth $B_c$ typically exceeding 1~MHz in indoor environments. The modulation features distinguishing Wi-Fi protocols (subcarrier spacing, guard intervals, pilot patterns) vary at or below the OFDM subcarrier spacing $\Delta\!f_{\text{sc}}$~\cite{ieee80211_2020}. This yields a \emph{scale separation}:
\vspace{-0.2cm}
\begin{equation}
\underbrace{B_c}_{\text{channel scale}} \;\gg\; \underbrace{\Delta\!f_{\text{sc}}}_{\text{modulation scale}}.
\label{eq:scale_separation}
\vspace{-0.2cm}
\end{equation}
Fig.~\ref{fig:system_model}c visualizes this scale separation on a real 
802.11ax capture.
This implies that any smoothing operation on the power spectrum with bandwidth 
$B_{\text{sm}}$ satisfying $\Delta\!f_{\text{sc}} \ll B_{\text{sm}} \ll B_c$ acts as a low-pass 
filter on the PSD. It tracks each transmitter's slow-varying channel envelope $|H_k(f)|^2$ within its occupied band while 
averaging out the fast-varying modulation structure of $S_k(f)$. Dividing the observed 
spectrum by this smoothed estimate therefore removes the channel-dependent component 
while preserving the protocol-discriminative features. For the non-OFDM 802.11b class, discrimination relies on time-domain DSSS chip structure, 
which whitening leaves intact even as it flattens the smooth DSSS spectral envelope. In the next section, we formalize 
this operation as \textit{spectral whitening}.

\vspace{-0.1cm}
\section{Proposed \ourmethod{} Framework}
\vspace{-0.1cm}
\label{sec:approach}

Fig.~\ref{fig:framework} provides an overview of the \ourmethod{} framework, which comprises four components:
\textbf{(i)}~spectral whitening for channel-invariant preprocessing,
\textbf{(ii)}~a physically grounded synthetic overlap mixer for training data generation,
\textbf{(iii)}~a compact encoder--decoder architecture with non-local (NL) attention, and
\textbf{(iv)}~knowledge distillation for edge deployment.

\vspace{-0.15cm}
\subsection{Spectral Whitening for Channel Removal}
\vspace{-0.1cm}
\label{subsec:spectral_whitening}

Motivated by the scale separation in Eq.~\eqref{eq:scale_separation}, we design a preprocessing step that removes the slow-varying channel envelope while preserving protocol-discriminative modulation structure. 
Given the digitized I/Q observation, we compute its $N$-point discrete Fourier transform (DFT) $Y[m]$ and estimate the channel-dominated 
PSD via circular moving-average smoothing:
\vspace{-0.3cm}
\begin{equation}
\hat{P}_{\text{ch}}[m] = \frac{1}{W} \sum_{j=0}^{W-1} \big|Y[(m{-}\lfloor W/2\rfloor{+}j) \bmod N]\big|^2,
\label{eq:psd_smooth}
\vspace{-0.2cm}
\end{equation}
where the window width $W$ (in bins) is chosen so that the smoothing bandwidth $B_{\text{sm}} = W f_s / N$, with $f_s$ the sampling rate, satisfies $\Delta\!f_{\text{sc}} \ll B_{\text{sm}} \ll B_c$. This condition ensures that $\hat{P}_{\text{ch}}[m]$ captures the channel envelope but averages out the modulation structure. A noise floor $\hat{P}_{\text{ch}}^{\,\text{f}}[m] = \max(\hat{P}_{\text{ch}}[m],\; P_{\max} \cdot 10^{\delta/10})$, where $P_{\max} = \max_m \hat{P}_{\text{ch}}[m]$, prevents amplification of deep fading nulls. The whitened spectrum is then obtained as:
\vspace{-0.35cm}
\begin{equation}
\tilde{Y}[m] = Y[m] \cdot \frac{\sqrt{\bar{P}}}{\sqrt{\hat{P}_{\text{ch}}^{\,\text{f}}[m]}},
\label{eq:whitening}
\vspace{-0.25cm}
\end{equation}
where $\bar{P} = \frac{1}{N}\sum_m \hat{P}_{\text{ch}}^{\,\text{f}}[m]$ normalizes total energy. Since $\hat{P}_{\text{ch}}^{\,\text{f}}[m]$ tracks the composite channel envelope at scales above $B_{\text{sm}}$ 
(each transmitter's $\alpha_k^2 |H_k(f)|^2$ within its occupied band), 
the whitened power spectrum satisfies $|\tilde{Y}[m]|^2 \propto |S_k(f)|^2$ within each transmitter's band. The channel response cancels, leaving the protocol-specific modulation structure intact. The same observation-derived procedure is applied at training and inference, requiring no channel knowledge. The estimate is derived entirely from the observation itself. The inverse DFT of $\tilde{Y}[m]$ yields the time-domain I/Q input to the classifier.



\vspace{-0.2cm}
\subsection{Synthetic Overlap Generation}
\vspace{-0.1cm}
\label{subsec:overlap_mixer}

Collecting real overlapping captures requires coordinating multiple transmitters across diverse channel conditions, an expensive process that limits dataset diversity. We instead generate synthetic overlapping signals from single-protocol OTA captures using a physically motivated signal chain. Given two single-protocol signals $s_1(t)$ and $s_2(t)$ extracted from non-overlapping OTA recordings, the synthetic composite is constructed as:
\vspace{-0.4cm}
\begin{equation}
y_{\text{syn}}(t) = \mathcal{R}\!\bigg\{\sum_{k=1}^{2} \alpha_k \big(g_k * s_k\big)(t)\, e^{j2\pi \Delta\!f_k t}\bigg\} + v(t),
\label{eq:synthetic_mix}
\vspace{-0.25cm}
\end{equation}
%
where $g_k(\tau)$ models per-transmitter multipath with independent tap gains and exponential power-delay profile, complemented by a random spectral tilt modeling antenna/cable frequency response. $\alpha_k$ captures power imbalance (near-far effect), and $\Delta\!f_k$ is a per-transmitter CFO. The receiver chain $\mathcal{R}\{\cdot\}$ is applied \emph{post-mixing}, since both signals pass through the same receiver, and includes an anti-aliasing low-pass filter, shared-oscillator phase noise (modeled as a Wiener process $\phi[n] = \phi[n{-}1] + \mathcal{N}(0, \sigma_\phi^2)$ applied identically to both signals), automatic gain control (AGC) with soft clipping, I/Q imbalance, and DC-offset injection (modeling LO leakage). All impairments are applied stochastically at the per-sample level. Thermal noise $v(t)$ is injected before the I/Q-imbalance and DC-offset stages, as in a physical front end. This distinction between per-transmitter impairments (applied \textit{before} summation) and shared receiver impairments (applied \textit{after}) is physically grounded and produces more realistic composite waveforms than simply adding two signals together.

Specifically, the receiver nonlinearity models front-end gain compression as a 
memoryless soft clipper $\mathcal{R}_{\text{nl}}(x) = \kappa \tanh(x / \kappa)$,
where $\kappa$ is proportional to the post-AGC RMS. 
This 
saturating function generates intermodulation products between co-channel signals, an 
effect absent from linear superposition, that is critical for realism when both 
transmitters operate at comparable power levels. Each signal is 
placed at a 
randomized 
center frequency within the wideband capture and occupies a randomized 
fraction of the observation window, producing diverse spectral and temporal overlap 
geometries that prevent the model from learning fixed positional priors.

\vspace{-0.25cm}
\subsection{Encoder--Decoder Architecture with Non-Local Attention}
\vspace{-0.1cm}
\label{subsec:architecture}

\ourmethod{} employs a U-Net-style~\cite{ronneberger2015u} encoder--decoder with a non-local attention module~\cite{wang2018non} at the bottleneck, followed by global average pooling (GAP) and a multi-label classification head. The encoder consists of four convolutional blocks with channel dimensions $C = [32, 64, 128, 256]$ and kernel sizes $[7, 5, 3, 3]$, where the larger kernels in early layers capture coarse temporal structure (e.g., the DSSS pattern of 802.11b) while smaller kernels in deeper layers capture fine-grained features (e.g., OFDM cyclic prefix boundaries). Each block applies two convolution--batch normalization--ReLU sequences followed by 2 times max-pooling, producing a bottleneck representation $\mathbf{F} \in \mathbb{R}^{256 \times 1024}$.
At the bottleneck, a non-local block computes self-attention over the 1{,}024 temporal positions. Given the bottleneck features $\mathbf{F}$, the block computes query $\boldsymbol{\Theta} = W_\theta \mathbf{F}$, key $\boldsymbol{\Phi} = W_\phi \mathbf{F}$, and value $\mathbf{G} = W_g \mathbf{F}$ projections, 
yielding an attention-weighted output:
\vspace{-0.15cm}
\begin{equation}
\mathbf{F}' = \mathbf{F} + W_z\, \mathbf{G}\, \text{softmax}\!\big(\boldsymbol{\Theta}^\top \boldsymbol{\Phi}\big)^{\!\top},
\label{eq:nonlocal}
\vspace{-0.25cm}
\end{equation}
where the residual connection preserves the original features. This mechanism captures long-range temporal dependencies.

The decoder mirrors the encoder through four upsampling blocks with skip connections that concatenate encoder features at matching resolutions, progressively restoring the temporal dimension. The final decoder output $\mathbf{D} \in \mathbb{R}^{16 \times N}$ is reduced via GAP to a 16-dimensional vector, passed through fully connected layers ($16 \to 64 \to 32 \to |\mathcal{P}|$) with ReLU activations and dropout, and mapped to per-protocol probabilities via independent sigmoid activations.
The rationale for a U-Net in a classification task, where encoder-only architectures are typical, is that the decoder, through skip connections, forces the network to maintain a complete multi-scale representation of the input signal rather than discarding fine-grained details during progressive downsampling. 

\vspace{-0.2cm}
\subsection{Knowledge Distillation for Edge Deployment}
\label{subsec:distillation}
\vspace{-0.1cm}

Where the deployment platform demands a lighter model, we compress the trained \ourmethod{} into four progressively smaller student architectures
(Table~\ref{tab:scaling}, Edge block).
The students follow a standard compression hierarchy that progressively trades architectural components for parameter efficiency: first reducing channel width, then removing the non-local block, then thinning the decoder to single-convolution blocks, and finally replacing the U-Net with a flat depthwise-separable classifier when the encoder--decoder overhead exceeds the available parameter budget. A challenge is that channel invariance, forged through the teacher's progressive training pipeline, does not transfer through standard logit-level distillation~\cite{hinton2015distilling} alone. Soft labels encode \emph{what} to predict but not \emph{which feature directions} the classifier should ignore. We therefore adopt a two-phase progressive distillation that mirrors the teacher's training trajectory in compressed form.

The first phase distills on synthetic overlap data from the pre-whitening teacher, building overlap decomposition capability without exposure to real channel conditions. The second phase fine-tunes on real OTA data using the whitened teacher, with three mechanisms that prevent \textit{catastrophic forgetting} of the first-phase representations. For the U-Net students, the earliest encoder block is frozen, preserving low-level signal processing learned on synthetic data.  The whitening schedule follows the same fixed-to-stochastic progression as the teacher, with proportionally milder stochastic ranges (Section~\ref{subsec:setup}), and a feature-alignment term is added to the distillation loss:
%
%
\vspace{-0.35cm}
\begin{multline}
\mathcal{L} = \lambda\,\mathcal{L}_{\text{BCE}} + (1{-}\lambda)\,T^2 \cdot \text{KL}\!\big(\boldsymbol{\sigma}(\mathbf{z}_T / T) \,\|\, \boldsymbol{\sigma}(\mathbf{z}_S / T)\big) \\
+ \;\beta\,\|\text{proj}(\mathbf{f}_S) - \mathbf{f}_T\|^2,
\label{eq:kd_loss}
\end{multline}
\vspace{-0.65cm}

\noindent where $\mathbf{z}_T$, $\mathbf{z}_S$ are teacher and student logits, $\boldsymbol{\sigma}$ denotes element-wise sigmoid, $T$ is the temperature, and $\lambda$ balances hard and soft targets. The third term matches the GAP-pooled bottleneck features $\mathbf{f}_T \in \mathbb{R}^{d_T}$ and $\mathbf{f}_S \in \mathbb{R}^{d_S}$ through a learned linear projection $\text{proj}: \mathbb{R}^{d_S} \to \mathbb{R}^{d_T}$, explicitly transferring the teacher's channel-invariant representation geometry to the student. Without this geometric constraint, the student is free to reach correct predictions via channel-dependent shortcuts that do \textit{not generalize} to unseen channel conditions.

\vspace{-0.25cm}
\section{Experimental Setup}
\label{subsec:setup}
\vspace{-0.1cm}

\noindent\textbf{Dataset.}~We use the publicly available OTA Wi-Fi dataset from~\cite{belgiovine2024tprime}. Each recording contains raw I/Q samples from two co-channel 20\,MHz Wi-Fi transmitters (802.11ax, b, g, n) captured by a wideband 62.5\,MHz SDR receiver, with varying degrees of spectral overlap. All sessions share the same indoor lab, hardware (Ettus USRP~X310 transmitters, AIR-T SDR receiver),
but were collected on separate days, resulting in different channel realizations. Table~\ref{tab:dataset} summarizes the partitioning. Sessions~S1 and S2 provide 6~unordered protocol pairs for training and validation (chronological 80/20 file-level split). Session~S3, collected on a different date, is reserved exclusively for held-out evaluation. It doubles the pair diversity to 12~ordered combinations (both transmitter orderings per pair), providing a more comprehensive test of generalization. The model has no direct or indirect exposure to S3 during training. All checkpoint and hyperparameter selection uses only the S1+S2 validation split.

\begin{table}[t]
\centering
\captionsetup{font=small}
\caption{Dataset partitioning. Session~S3 is held out for channel robustness evaluation and is never used during training or validation.
}
\vspace{-0.2cm}
\label{tab:dataset}
\scriptsize
\setlength{\tabcolsep}{3pt}
\renewcommand{\arraystretch}{1.1}
\begin{tabular}{@{}lcccl@{}}
\toprule
\textbf{Session} & \textbf{Overlap\textsuperscript{$\dagger$}} & \textbf{Pairs} & \textbf{Files} & \textbf{Role} \\
\midrule
S1 & 0/25/50\% & 6 & 4{,}800 & Train/Val (80/20) \\
S2 & 25/50\%   & 6 & 2{,}400 & Train/Val (80/20) \\
\midrule
S3 & 25\% & 12 & 2{,}400 & \textbf{Held-out test} \\
S3 & 50\% & 12 & 2{,}400 & \textbf{Held-out test} \\

\bottomrule
\multicolumn{5}{@{}l}{\rule{0pt}{2.2ex}\scriptsize \textsuperscript{$\dagger$}Spectral overlap between the two co-channel 
transmitters.}
\end{tabular}
\vspace{-0.75cm}
\end{table}

\noindent\textbf{Training Pipeline.}~Rather than training end-to-end on limited OTA data, we employ a progressive transfer strategy that sequentially introduces task complexity, from clean single-protocol waveforms to real overlapping captures with channel whitening. Each phase initializes from the best checkpoint of the preceding phase, so representational capacity accumulates across the full pipeline. Table~\ref{tab:pipeline} summarizes all five phases.

Phases~A--B establish protocol discrimination on synthetic 20\,MHz waveforms augmented with a nine-impairment wireless channel model, then transfer to real OTA captures at 31.25\,MHz. Phases~C--D introduce multi-label overlap classification. Phase~C generates 50K training samples on-the-fly via the overlap mixer (Eq.~\ref{eq:synthetic_mix}), of which half are overlapping pairs and half single-protocol to retain single-signal discrimination, with randomized temporal placement, power dominance, adjacent-channel frequency shifts, and receiver nonlinearity (AGC + soft clipping). Phase~D scales to 500K pre-generated samples with the same composition at the target 62.5\,MHz bandwidth. 
All phases use AdamW with cosine-annealed learning rate, early stopping, and input length $N{=}16{,}384$ I/Q samples (262\,$\mu$s at 62.5\,MHz sampling rate), sufficient to capture at least one complete Wi-Fi preamble across all four protocol versions, and RMS-normalized to unit power.

\begin{table}[t]
\centering
\captionsetup{font=small}
\caption{Progressive training pipeline. Each phase inherits the preceding checkpoint. Phases~A--D build protocol and overlap representations from synthetic or single-protocol data. Phase~E introduces channel robustness via spectral whitening on real OTA captures.}
\vspace{-0.2cm}
\label{tab:pipeline}
\scriptsize
\setlength{\tabcolsep}{2pt}
\renewcommand{\arraystretch}{1.15}
\begin{tabular}{@{}cllllc@{}}
\toprule
\textbf{Ph.} & \textbf{Data Source} & \textbf{Task} & \textbf{Key Objective} & \textbf{Augmentation} & \textbf{Ep.} \\
\midrule
A & Synthetic, 20\,MHz & Single & Protocol discrim. & Full channel$^{a}$ & 100 \\
B & OTA single, 31.25\,MHz & Single & Syn$\to$real transfer & Full channel$^{a}$ & 100 \\
C & Synth.\ overlap, 31.25\,MHz & Multi & Overlap separation & TX/RX + nonlin.$^{b}$ & 100 \\
D & Synth.\ overlap, 62.5\,MHz & Multi & BW scaling & TX/RX + nonlin.$^{b}$ & 70 \\
E & OTA overlap (S1+S2) & Multi & Channel robustness & Light + whiten.$^{c}$ & 200 \\
\bottomrule
\multicolumn{6}{@{}p{0.95\columnwidth}}{\rule{0pt}{2.2ex}\scriptsize
$^{a}$Phase rotation, path loss, Rician/Rayleigh fading, CFO, AWGN ($-$5---30\,dB), I/Q imbalance, phase noise, DC offset, time shift.
$^{b}$Per-TX: phase rotation, time shift, CFO (Phase~D adds multipath fading and spectral tilt); per-RX: AA filter, I/Q imbalance, DC offset, AWGN (Phase~D adds common-LO phase noise); nonlinearity: AGC + soft clipping.
$^{c}$Light augmentation (phase rotation, time shift) 
+ spectral whitening with progressive schedule.}
\end{tabular}
\vspace{-0.65cm}
\end{table}

Phase~E fine-tunes on 5{,}760 real OTA overlap captures from S1+S2 with \textit{spectral whitening} as preprocessing. 
Since each capture is longer than $N$, a random $N$-sample window is extracted per capture in each epoch, so the model sees different signal segments across epochs without inflating the dataset with correlated crops.
The first two encoder blocks are frozen to preserve overlap representations. Dropout is raised to 0.4 throughout the network to regularize fine-tuning.
Training exclusively with fixed whitening parameters risks overfitting to one specific residual spectral shape, while fully stochastic whitening from the start destabilizes early learning before protocol features have consolidated. Whitening therefore follows a progressive schedule: fixed parameters ($W{=}256$, $\delta{=}{-}30$\,dB) for epochs 1--80 allow stable feature consolidation, a linear transition over epochs 80--140 gradually introduces parameter variation, and predominantly stochastic whitening ($W{\in}[64,1024]$, $\delta{\in}[-40,-20]$\,dB, applied to 80\% of samples) with synthetic channel injection thereafter teaches invariance to diverse whitening residuals. At inference, only fixed whitening is applied. 


\noindent\textbf{Performance Metrics.}~We report two multi-label accuracy metrics. \emph{Exact-match accuracy} (EM) requires all $|\mathcal{P}|{=}4$ protocol labels to be correct simultaneously:
\vspace{-0.25cm}
\begin{equation}
\text{EM} = \frac{1}{M}\sum_{i=1}^{M} \mathds{1}\!\bigl[\hat{\mathbf{l}}_i = \mathbf{l}_i\bigr],
\label{eq:em}
\vspace{-0.2cm}
\end{equation}
where $\hat{\mathbf{l}}_i, \mathbf{l}_i \in \{0,1\}^{|\mathcal{P}|}$ and $M$ is the number of evaluation samples. \emph{Hamming accuracy} (HA) measures the fraction of individually correct labels:
\vspace{-0.25cm}
\begin{equation}
\text{HA} = \frac{1}{M|\mathcal{P}|}\sum_{i=1}^{M}\sum_{j=1}^{|\mathcal{P}|} \mathds{1}\!\bigl[\hat{l}_{ij} = l_{ij}\bigr].
\label{eq:ha}
\vspace{-0.2cm}
\end{equation}

\noindent\textbf{Hardware characteristics.}~Training is conducted on a single NVIDIA RTX PRO 6000 (96\,GB GDDR7). Edge inference benchmarks use an NVIDIA Jetson Orin Nano (8\,GB LPDDR5, 1024-core Ampere GPU with 32 tensor cores),
representative of deployable spectrum-monitoring hardware.

\vspace{-0.15cm}
\section{Results and Analysis}
\label{subsec:results}
\vspace{-0.1cm}


\noindent\textbf{Preamble-Based Detection as Baseline.}~We first quantify what standard-compliant PHY preamble detection~\cite{ieee80211_2020} achieves on the same 262\,$\mu$s snapshots the classifier receives, granting it \emph{oracle knowledge of both transmitter center frequencies}. 
Family-level preamble detection (Barker correlation for DSSS, L-STF autocorrelation~\cite{schmidl1997robust} for OFDM) recognizes both constituent families in only 50.2\% of non-overlapping captures and 17.7--18.0\% under overlap, and cannot distinguish the OFDM generations at all. Protocol-level discrimination via L-STF synchronization, L-LTF channel estimation, and RL-SIG/HT-SIG inspection identifies both protocols in at most 3.8\% of captures even without spectral overlap, since 802.11b exposes no SIG field, and in at most 1.0\% under overlap, while a channelized receiver chain with the standard's mutually exclusive DSSS/OFDM detection logic never identifies both. 
Joint protocol identification under coexistence is therefore beyond conventional PHY preamble detection even in its most \textit{favorable condition}.

\begin{figure}[t]
\centering
\includegraphics[width=\columnwidth]{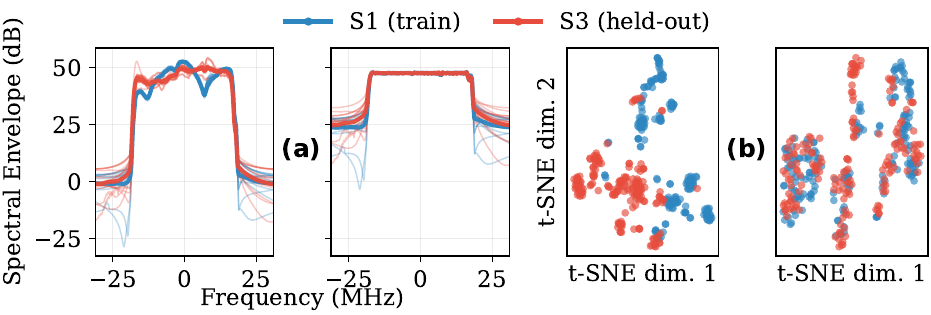}
\captionsetup{font=small}
\vspace{-0.55cm}
\caption{Channel distribution shift (ax+b, OV25). (a)~Spectral envelope before/after whitening: S1--S3 divergence drops from 3.45\,dB to 0.063\,dB. (b)~t-SNE of bottleneck features: session separability drops from 88.9\% to 59.4\% after whitening.}
\label{fig:channel_shift}
\vspace{-0.85cm}
\end{figure}

\noindent\textbf{Pipeline Validation (Phases A--D).}~Before addressing channel robustness, we verify that each pipeline stage delivers its intended capability. After Phase~B, the model achieves 99.99\% accuracy on held-out OTA single-protocol captures at 31.25\,MHz, confirming that the architecture learns protocol-discriminative features from raw I/Q alone.
Phases~C--D then extend these features to multi-label overlap classification, reaching 87.2\% EM accuracy on synthetic overlap at 62.5\,MHz. Together, Phases~A--D establish a representation that reliably separates protocols and handles spectral overlap, the prerequisite for Phase~E, which must only learn to be \emph{invariant} to channel variation rather than learn 
from scratch.

\begin{figure*}[t]
\centering
\includegraphics[width=0.75\textwidth]{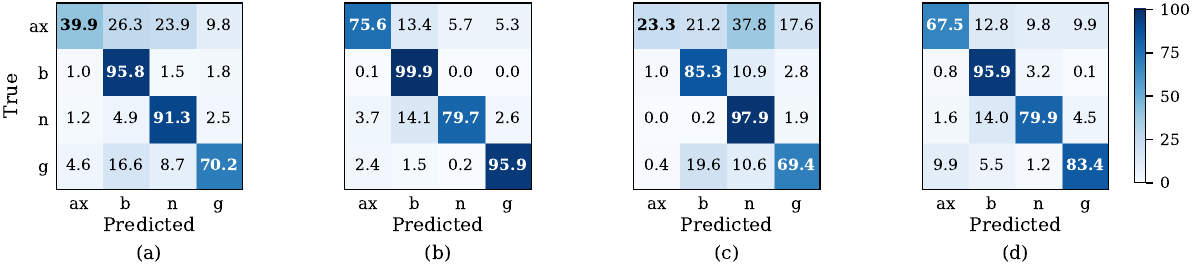}
\captionsetup{font=small}
\vspace{-0.2cm}
\caption{Multi-label confusion matrices on held-out S3 (2\,400 samples each). (a,\,c)~Baseline, (b,\,d)~\ourmethod{} at OV25 and OV50. 
}
\label{fig:confusion}
\vspace{-0.6cm}
\end{figure*}

\noindent\textbf{Channel Distribution Shift.}~Fig.~\ref{fig:channel_shift}a visualizes the spectral envelope (smoothed with the same 256-bin window used by whitening) for 15 individual captures of the same protocol pair (ax+b, OV25) from S1 and S3. Before whitening, the two sessions exhibit distinct envelope shapes with a mean in-band divergence of 3.45\,dB and per-capture dynamic ranges of 7--18\,dB, reflecting different multipath channel realizations. After whitening, all envelopes collapse to a near-flat profile with only 0.063\,dB mean divergence (55 times reduction). Fig.~\ref{fig:channel_shift}b confirms this at the feature level. In the baseline model, bottleneck features from S1 and S3 cluster by session (88.9\% 5-NN session separability), mirroring the channel dominance previously shown for RF fingerprinting~\cite{al2020exposing}.
After whitening, the features interleave with 59.4\% 5-NN separability, approaching the 50\% random-guess level, which indicates that the nearest-neighbor classifier can no longer reliably distinguish which session a sample originated from and confirms that the learned representations are largely channel-invariant.

\noindent\textbf{Channel Robustness Evaluation.}
Table~\ref{tab:channel_robust} presents the central evaluation. The \emph{in-distribution} column reports EM on the chronological held-out split from S1+S2 (i.e., same capture sessions as training). The \emph{held-out} columns report EM on the entirely unseen S3 at 25\% (OV25) and 50\% (OV50) 
overlap. 

\begin{table}[t]
\centering
\captionsetup{font=small}
\caption{Channel robustness evaluation. All Phase~E methods share the same Phase~A$\to$D backbone. Published baselines are retrained 
on S1+S2. \emph{In-Dist.}\ = S1+S2 held-out; \emph{Held-Out} = unseen S3 at 62.5\,MHz native rate.}
\vspace{-0.2cm}
\label{tab:channel_robust}
\scriptsize
\setlength{\tabcolsep}{2pt}
\renewcommand{\arraystretch}{1.15}
\begin{tabular}{@{}lcc|cc|cc|c@{}}
\toprule
& \multicolumn{2}{c|}{\textbf{In-Dist.}} & \multicolumn{2}{c|}{\textbf{Held-Out OV25}} & \multicolumn{2}{c|}{\textbf{Held-Out OV50}} & \\
\cmidrule(lr){2-3} \cmidrule(lr){4-5} \cmidrule(lr){6-7}
\textbf{Phase E Method} & \textbf{EM} & \textbf{HA} & \textbf{EM} & \textbf{HA} & \textbf{EM} & \textbf{HA} & $\boldsymbol{\Delta}$\textbf{OV25} \\
\midrule
Baseline (no whitening) & \textbf{87.6} & \textbf{94.4} & 47.6 & 75.0 & 38.3 & 72.5 & --- \\
\quad + DANN~\cite{ganin2016dann} & 81.5 & 92.1 & 50.0 & 76.8 & 42.7 & 74.1 & +2.4 \\
\quad + Reptile~\cite{nichol2018reptile} & 40.3 & 74.5 & 38.8 & 76.9 & 23.7 & 76.9 & $-$8.8 \\
\textbf{\ourmethod{} (whitening)}$^*$ & 81.6 & 92.7 & \textbf{73.6} & \textbf{88.1} & \textbf{60.5} & \textbf{82.3} & \textbf{+26.0} \\
\midrule
\multicolumn{8}{@{}l}{\textit{Published baselines:}} \\
T-PRIME~\cite{belgiovine2024tprime}$^\dagger$ & 69.4 & 87.5 & 32.2 & 69.2 & 23.5 & 62.6 & $-$15.4 \\
ResNet~\cite{oshea2018over}$^\ddagger$ & 79.3 & 89.9 & 28.2 & 61.5 & 28.2 & 60.5 & $-$19.4 \\
Stitching U-Net~\cite{uvaydov2024stitching}$^\S$ & 68.5 & 89.5 & 19.8 & 60.4 & 11.6 & 56.3 & $-$27.8 \\
JDM~\cite{xie2024joint_detection_amc}$^\P$ & 64.2 & 82.4 & 22.2 & 59.1 & 7.9 & 53.8 & $-$25.4 \\
CV-TRN~\cite{li2024complex}$^\|$ & 81.5 & 91.5 & 22.0 & 58.2 & 5.0 & 51.2 & $-$25.6 \\
\bottomrule
\multicolumn{8}{@{}l}{\rule{0pt}{2.2ex}\scriptsize Params of retrained models: $^*$889K; $^\dagger$6.8M; $^\ddagger$101K; $^\S$13.6M; $^\P$321K; $^\|$45.7K.}\\
\end{tabular}
\vspace{-0.9cm}
\end{table}

Phase~E methods
in Table~\ref{tab:channel_robust} share the same Phase~A--D backbone, which was trained with aggressive channel augmentation (random multipath, CFO, and AWGN). The baseline row therefore already reflects the best that channel augmentation alone can achieve. The +26.0\,pp gain from whitening is attributable to the whitening step itself. The in-distribution--held-out gap narrows from 40.0\,pp (baseline) to 8.0\,pp (\ourmethod{}).
This confirms the hypothesis from Section~\ref{subsec:scale_separation}. The channel robustness gap is fundamentally a \emph{channel envelope} problem, and removing it at the signal level is the most direct remedy. Simpler normalizations lack the adaptive, per-signal envelope estimation that spectral whitening provides, and do not exploit the scale separation between channel coherence bandwidth and subcarrier spacing. Standard instance normalization~\cite{ulyanov2016instance} 
yields 43.2\% EM at OV25, indistinguishable from removing whitening entirely (43.1\%), because time-domain statistics cannot address frequency-selective fading. Per-signal spectral flattening 
is worse still, collapsing to 23.5\% EM at OV25 (HA: 69.4\%) as it destroys protocol-discriminative modulation structure along with the channel envelope. The trade-off in \ourmethod{} is a 6.0\,pp in-distribution EM decrease, 
because whitening removes channel-correlated cues that happen to aid classification when training and test conditions coincide. This cost is modest relative to the 26.0\,pp held-out recovery and reflects the removal of spurious features that would not generalize.

\begin{figure}[t]
\centering
\includegraphics[width=0.8\columnwidth]{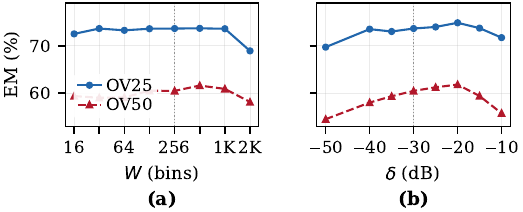}
\captionsetup{font=small}
\vspace{-0.2cm}
\caption{Sensitivity of whitening parameters on held-out S3. (a)~Smoothing window $W$; the corresponding smoothing bandwidth $B_{\text{sm}} = W f_s / N$ ranges from 0.06 to 7.8\,MHz. (b)~Noise floor $\delta$. Dotted lines mark the defaults ($W{=}256$, $\delta{=}{-30}$\,dB). 
}
\label{fig:sensitivity}
\vspace{-0.75cm}
\end{figure}

Fig.~\ref{fig:sensitivity} examines the sensitivity of whitening to its two hyperparameters. The smoothing window  is varied across $W \in [16, 1024]$
with less than 1.2\,pp variation in OV25 EM (72.5--73.7\%) and 2.6\,pp in OV50 (59.0--61.6\%), while HA remains at or above 87.7\% and 81.8\% respectively throughout the entire sweep. This broad plateau confirms that the scale separation condition $\Delta\!f_{\text{sc}} \ll B_{\text{sm}} \ll B_c$ is easily satisfied in practice. Performance degrades only at $W = 2048$, where the smoothing bandwidth ($B_{\text{sm}} = 7.8$\,MHz) approaches the channel coherence bandwidth and begins averaging out protocol structure (OV25 EM drops to 68.9\%). The noise floor $\delta$ exhibits a similarly broad optimum over $[-40, -15]$\,dB (OV25 EM varies only 1.8\,pp). At the extremes, $\delta = -50$\,dB fails to protect fading nulls (OV25: 69.7\%, OV50: 54.5\%) while $\delta = -10$\,dB over-suppresses useful spectral content (OV25: 71.7\%, OV50: 55.8\%). The default parameters ($W{=}256$, $\delta{=}{-30}$\,dB) lie within both plateaus, requiring no dataset-specific tuning.

\begin{table*}[t]
\centering
\captionsetup{font=small}
\caption{Scaling and edge deployment. Input and architecture blocks use the full Phase~A$\to$E pipeline; Edge models are distilled from the \textit{tiny} teacher (Section~\ref{subsec:distillation}). Latency/energy on Jetson Orin Nano (15\,W); GPU latency on RTX PRO 6000. 
}
\vspace{-0.2cm}
\label{tab:scaling}
\scriptsize
\setlength{\tabcolsep}{3pt}
\renewcommand{\arraystretch}{1.12}
\begin{tabular}{@{}ll l r r r r cc cc cc cc c@{}}
\toprule
& & & & & & & \multicolumn{2}{c}{\textbf{In-Dist.}} & \multicolumn{2}{c}{\textbf{Held-Out OV25}} & \multicolumn{2}{c}{\textbf{Held-Out OV50}} & \multicolumn{2}{c}{\textbf{Latency (ms)}} & \textbf{Energy} \\
\cmidrule(lr){8-9} \cmidrule(lr){10-11} \cmidrule(lr){12-13} \cmidrule(lr){14-15}
& \textbf{Model} & \textbf{Type$^{*}$} & \textbf{$N$} & \textbf{Params} & \textbf{GFLOPs} & \textbf{Mem} & \textbf{EM} & \textbf{HA} & \textbf{EM} & \textbf{HA} & \textbf{EM} & \textbf{HA} & \textbf{RTX} & \textbf{Jetson} & \textbf{} \\
& & & & & & \textbf{(MB)} & \textbf{(\%)} & \textbf{(\%)} & \textbf{(\%)} & \textbf{(\%)} & \textbf{(\%)} & \textbf{(\%)} & & & \textbf{(mJ)} \\
\midrule
\multirow{3}{*}{\rotatebox[origin=c]{90}{\scriptsize\textbf{Input}}}
& Tiny & 4L+NL & 8{,}192 & 889K & 2.54 & 14.8 & 61.0 & 86.9 & 56.5 & 84.1 & 38.2 & 77.2 & 1.02 & 9.9 & 137.2 \\
& \cellcolor{blue!6}\textbf{Tiny (ours)} & \cellcolor{blue!6}\textbf{4L+NL} & \cellcolor{blue!6}\textbf{16{,}384} & \cellcolor{blue!6}\textbf{889K} & \cellcolor{blue!6}\textbf{5.08} & \cellcolor{blue!6}\textbf{24.3} & \cellcolor{blue!6}\textbf{81.6} & \cellcolor{blue!6}\textbf{92.7} & \cellcolor{blue!6}\textbf{73.6} & \cellcolor{blue!6}\textbf{88.1} & \cellcolor{blue!6}\textbf{60.5} & \cellcolor{blue!6}\textbf{82.3} & \cellcolor{blue!6}\textbf{1.03} & \cellcolor{blue!6}\textbf{11.7} & \cellcolor{blue!6}\textbf{205.6} \\
& Tiny & 4L+NL & 32{,}768 & 889K & 10.16 & 59.1 & 63.3 & 88.4 & 61.0 & 85.2 & 44.2 & 78.5 & 1.04 & 24.9 & 445.7 \\
\midrule
\multirow{3}{*}{\rotatebox[origin=c]{90}{\scriptsize\textbf{Arch.}}}
& Small & 5L+NL & 16{,}384 & 3.5M & 9.79 & 45.3 & 66.5 & 89.0 & 63.1 & 83.0 & 45.3 & 76.7 & 1.26 & 16.4 & 285.7 \\
& Medium & 5L+NL & 16{,}384 & 13.8M & 39.05 & 124.3 & 78.2 & 92.0 & 72.2 & 86.7 & 53.4 & 78.7 & 1.43 & 41.7 & 783.2 \\
& Large & 6L+NL & 16{,}384 & 29.9M & 57.04 & 188.9 & 76.2 & 92.5 & 89.9 & 95.4 & 66.5 & 85.6 & 1.91 & 62.5 & 1168.6 \\
\midrule
\multirow{4}{*}{\rotatebox[origin=c]{90}{\scriptsize\textbf{Edge}}}
& Edge-L & 4L+NL & 16{,}384 & 604K & 4.98 & 25.5 & 71.0 & 89.8 & 61.8 & 82.4 & 47.1 & 76.1 & 0.90 & 13.4 & 233.5 \\
& Edge-M & 4L & 16{,}384 & 322K & 2.84 & 19.7 & 61.6 & 87.5 & 50.6 & 77.5 & 39.8 & 75.0 & 0.71 & 10.0 & 169.9 \\
& Edge-S & 4L-thin & 16{,}384 & 171K & 1.27 & 12.3 & 61.3 & 87.1 & 49.2 & 78.6 & 39.9 & 75.9 & 0.62 & 6.8 & 102.1 \\
& \cellcolor{red!6}Edge-T & \cellcolor{red!6}Flat-DW & \cellcolor{red!6}16{,}384 & \cellcolor{red!6}10K & \cellcolor{red!6}0.03 & \cellcolor{red!6}1.6 & \cellcolor{red!6}36.7 & \cellcolor{red!6}80.0 & \cellcolor{red!6}34.6 & \cellcolor{red!6}74.9 & \cellcolor{red!6}26.1 & \cellcolor{red!6}69.9 & \cellcolor{red!6}0.30 & \cellcolor{red!6}3.1 & \cellcolor{red!6}33.7 \\
\bottomrule
\multicolumn{16}{@{}l}{\rule{0pt}{2.2ex}\scriptsize $^{*}$\textit{n}L = \textit{n} encoder levels; NL = non-local attention; thin = narrowed channels with single-convolution decoder blocks.}\\
\multicolumn{16}{@{}l}{\scriptsize \phantom{$^{*}$}Flat-DW = MobileNet-style depthwise-separable classifier (no U-Net).}
\end{tabular}
\vspace{-0.7cm}
\end{table*}

\noindent\textbf{Domain-Adaptation Baselines.}~We include DANN~\cite{ganin2016dann} and Reptile~\cite{nichol2018reptile} as reference points (Table~\ref{tab:channel_robust}) for the two dominant adaptation paradigms (adversarial and meta-learning), while acknowledging that neither is designed for the low-diversity regime of two training sessions. DANN yields only +2.4\,pp on OV25 at the cost of $-$6.1\,pp in-distribution, because the domain discriminator lacks sufficient session diversity to learn meaningful invariances. Reptile degrades exact-match accuracy across all conditions, as the task-alternation regime destabilizes representations acquired over Phases~A--D. These results confirm that \emph{learning-based} domain adaptation requires substantially more diverse training domains than are typically available in OTA capture campaigns, motivating the signal-level approach taken by \textit{spectral whitening}.


To contextualize these gains we retrained five published classifiers (Table~\ref{tab:channel_robust}) on the same overlap data (S1+S2),
and evaluated them on the unseen S3 under the identical protocol: T-PRIME~\cite{belgiovine2024tprime}, the ResNet of~\cite{oshea2018over}, the spectrum-segmentation U-Net of~\cite{uvaydov2024stitching} (the architecture family our classifier builds on) adapted to multi-label classification, 
JDM~\cite{xie2024joint_detection_amc} granted oracle transmitter frequencies in place of its detection stage, and the complex-valued transformer (CV-TRN)~\cite{li2024complex}. Every one learns the task when channels match, reaching 64.2--81.5\% in-distribution EM, and every one collapses on the unseen session to 19.8--32.2\% EM at OV25 and 5.0--28.2\% at OV50. CV-TRN is the starkest case, matching \ourmethod{}'s in-distribution accuracy (81.5\% vs.\ 81.6\%) yet retaining only 22.0\% held-out where \ourmethod{} retains 73.6\%, so neither in-distribution accuracy, parameter count (45.7K--13.6M), nor architecture family predicts channel robustness. \ourmethod{} more than doubles the best baseline held-out accuracy (T-PRIME, 32.2\%) while using 7.7 times fewer parameters. 

\begin{figure}[t]
\vspace{-0.15cm}
\centering
\includegraphics[width=0.7\columnwidth]{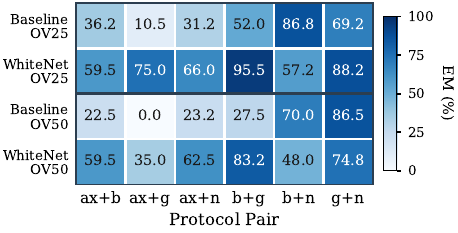}
\captionsetup{font=small}
\vspace{-0.2cm}
\caption{Per-pair EM accuracy (\%) across four conditions. OV25 (top) and OV50 (bottom). The ax+g pair benefits most from whitening at OV25; b+n is the only pair that degrades
at OV25/OV50.}
\label{fig:per_pair}
\vspace{-0.75cm}
\end{figure}

\noindent\textbf{Per-Protocol Breakdown.}~Fig.~\ref{fig:confusion} and Fig.~\ref{fig:per_pair} reveal \emph{where} whitening helps and where challenges remain. In the baseline at OV25 (Fig.~\ref{fig:confusion}a), \textit{802.11ax recall is only 39.9\%} because the channel envelope masks the HE-SIG-A preamble that distinguishes ax from other OFDM variants. After whitening (b), \textit{ax recall rises to 75.6\%} (+35.7\,pp). Similarly, 802.11g improves from 70.2\% to 95.9\%, while 802.11b remains near-perfect (95.8\%\,$\to$\,99.9\%). The one protocol that loses recall is \textit{802.11n} (91.3\%\,$\to$\,79.7\%), because n and g share nearly identical OFDM structure and the baseline exploits channel-correlated cues that do not generalize. However, 802.11n precision improves substantially (64.6\%\,$\to$\,88.8\%, F1\,=\,84.0\%), reflecting better calibration rather than systematic failure. At OV50, \ourmethod{} retains 67.5\% ax and 83.4\% g recall where the baseline collapses to 23.3\% and 69.4\%, respectively.

Fig.~\ref{fig:per_pair} disaggregates by \textit{protocol pair}. The ax+g pair benefits most (+64.5\,pp at OV25, from 10.5\% to 75.0\%), because both protocols rely on channel-sensitive preamble fields that whitening recovers. The b+g pair follows (+43.5\,pp), as the robust DSSS signature of 802.11b combines with improved g classification. The only pair that degrades is b+n 
($-$29.6\,pp at OV25), consistent with the per-class n recall decrease. At OV50, whitening maintains large gains for ax+g (+35.0\,pp) and b+g (+55.7\,pp), though absolute EM is lower due to increased spectral congestion.

\begin{figure*}[t]
\centering
\includegraphics[width=0.75\textwidth]{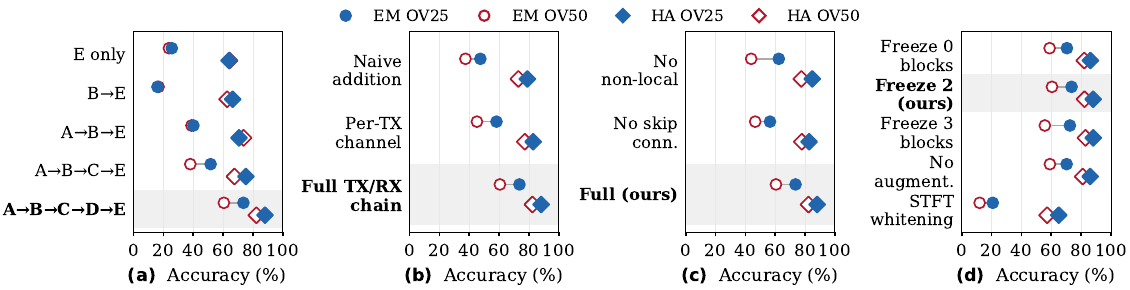}
\captionsetup{font=small}
\vspace{-0.2cm}
\caption{Ablation study reporting held-out EM and HA for OV25 and OV50. (a)~Progressive pipeline phases. (b)~Synthetic overlap generation strategy. (c)~Architecture components. (d)~Fine-tuning configuration. The highlighted row marks our full system.}
\label{fig:ablation}
\vspace{-0.65cm}
\end{figure*}

\noindent\textbf{Input Length and Architecture Scaling.}~Table~\ref{tab:scaling} reports the effect of observation window length and model capacity on classification accuracy, computational cost, and edge-device efficiency. Doubling the observation window from $N{=}8{,}192$ (131\,$\mu$s) to $N{=}16{,}384$ (262\,$\mu$s) lifts OV25 EM from 56.5\% to 73.6\%, an improvement of 17.1\,pp, because the longer window captures additional preamble repetitions and cyclic prefix structure that disambiguate overlapping protocols. Jetson latency increases only marginally (9.9 to 11.7\,ms) because fixed per-inference overheads dominate at these input sizes, although compute doubles (2.54 to 5.08\,GFLOPs). Extending to $N{=}32{,}768$ recovers some of this gap (61.0\% OV25 EM) but does not surpass $N{=}16{,}384$, while Jetson energy more than doubles (206 to 446\,mJ), indicating diminishing returns beyond 262\,$\mu$s observation windows for this protocol set.

For architecture scaling, the Tiny model (889K, four encoder levels) provides the strongest accuracy-per-compute trade-off. The Large model (29.9M) achieves 89.9\% OV25 EM, the highest in the table, but requires eleven times the compute (57 vs.\ 5.1\,GFLOPs) and nearly six times the energy on the Jetson (1169 vs.\ 206\,mJ). The Small and Medium models (3.5M and 13.8M) underperform the Tiny model on held-out OV25 EM despite four to fifteen times more parameters, while the Large model recovers and even exceeds its own in-distribution accuracy on held-out data (89.9\% vs.\ 76.2\% EM). A width-matched 3.5M control with the same 4-level topology as Tiny, trained through the identical pipeline, reaches only 59.0\% OV25 EM, indicating that the \textit{non-monotonic pattern} stems from optimization dynamics of mid-sized models under the 5{,}760-sample Phase~E budget rather than from encoder depth or simple capacity overfitting. 

\noindent\textbf{Edge Deployment.}~The bottom block of Table~\ref{tab:scaling} reports knowledge-distilled students compressed from the whitened teacher via the two-phase progressive distillation described in Section~\ref{subsec:distillation}. The Edge-L student retains 84\% of the teacher's held-out accuracy, reaching 61.8\% OV25 EM and 82.4\% Hamming accuracy on channel conditions never seen during training. Edge-M preserves the multi-scale decoder while removing the non-local block, retaining 50.6\% OV25 EM (77.5\% HA). Edge-S matches this accuracy with roughly half the parameters by narrowing the channel widths and thinning each decoder stage to a single convolution (49.2\% EM, 78.6\% HA). Both demonstrate that even at up to five-fold compression relative to the teacher, the models correctly identify at least one active protocol in the majority of overlap cases. The Edge-T model reaches 34.6\% OV25 EM (74.9\% HA) at 3.1\,ms and 33.7\,mJ on the Jetson, making it suitable for coarse spectrum awareness on 
power-constrained platforms.

\noindent\textbf{Ablation Study.}~Fig.~\ref{fig:ablation} isolates the contribution of each design decision along four axes, reporting held-out OV25/OV50 EM and HA for the Tiny architecture ($N{=}16{,}384$). Each ablation modifies a single factor while holding all others constant.

\noindent \tikz[baseline=(char.base)]\node[shape=circle, fill=black, inner sep=0.7pt, text=white] (char) {1}; \textbf{\textit{{Pipeline phases.}}}~Training directly on real OTA overlap (``E only'') achieves just 25.6\% OV25 EM, confirming that 5{,}760 captures cannot teach multi-label classification from scratch. Skipping from single-protocol pre-training to real overlap (B$\to$E) is worse still: OV25 EM drops to 16.0\%, \emph{below} the from-scratch baseline, yet HA remains at 66.4\%, indicating the model recognizes protocols individually but the abrupt domain shift destabilizes overlap decomposition. Adding augmentation-hardened Phase~A (A$\to$B$\to$E) recovers EM to 40.1\%. Introducing synthetic overlap at 31.25\,MHz (A$\to$B$\to$C$\to$E) lifts it to 51.6\%, though OV50 stalls at 38.0\% due to bandwidth mismatch. Only the full pipeline
reaches 73.6\%/60.5\% OV25/OV50 EM.
The monotonic improvement as phases are added (16.0\,$\to$\,40.1\,$\to$\,51.6\,$\to$\,73.6\% OV25 EM) confirms that each phase addresses a \textit{failure mode} later stages alone cannot recover from.

\noindent \tikz[baseline=(char.base)]\node[shape=circle, fill=black, inner sep=0.7pt, text=white] (char) {2};\textbf{\textit{{~Overlap generation.}}}~Replacing the physically grounded mixer (Eq.~\ref{eq:synthetic_mix}) with naive signal addition caps OV25 EM at 47.4\% (OV50: 37.4\%), yet HA already reaches 78.9\%, showing the model learns coarse signatures but cannot resolve co-channel boundaries. Adding per-TX channel models lifts OV25 EM to 58.2\% by forcing the network to disentangle overlapping envelopes rather than memorize fixed power ratios. The full TX/RX chain, including shared AGC, phase noise, and soft clipping, yields 73.6\%/60.5\% OV25/OV50 EM. The 15.4\,pp jump from per-TX to full chain confirms that modeling \textit{shared receiver distortions} is as important as modeling independent transmitter channels.

\noindent \tikz[baseline=(char.base)]\node[shape=circle, fill=black, inner sep=0.7pt, text=white] (char) {3}; \textbf{\textit{{Architecture.}}}~Removing the \textit{NL attention} block drops OV25 EM from 73.6\% to 62.4\% and OV50 from 60.5\% to 43.9\%. Without global self-attention the model cannot relate temporally distant preamble fields (e.g., L-STF to HE-SIG-A, separated by 1600 samples) that local convolutions miss. This component adds only 132K parameters yet delivers the largest per-component gain under severe overlap (16.6\,pp OV50 EM). Removing \textit{skip connections} is more damaging at moderate overlap (OV25 EM collapses to 56.5\%), as the decoder loses fine-grained modulation structure from early encoder layers. Interestingly, OV50 EM under no-skip (46.6\%) slightly exceeds the no-NL variant (43.9\%), suggesting that under severe overlap the global receptive field matters more than multi-scale detail. Together, both components bring the model to 889K parameters and 5.08\,GFLOPs. Skip connections alone account for a 21\% compute overhead and the NL block a further 6\%.

\noindent \tikz[baseline=(char.base)]\node[shape=circle, fill=black, inner sep=0.7pt, text=white] (char) {4};\textbf{\textit{{~Fine-tuning.}}}~Full fine-tuning (freeze~0) erodes overlap decomposition learned in Phases~C--D, dropping OV25 EM to 70.5\%. Over-constraining (freeze~3) retains OV25 HA at 88.1\% but sacrifices OV50 EM to 55.7\%. Freezing the first two blocks balances both at 73.6\%/60.5\%. 
Removing the light augmentation applied to real captures during Phase~E (random phase rotation, sub-window time shifts)
reduces OV25 EM to 70.3\%, confirming that augmentation diversity remains beneficial even in the final fine-tuning stage. 
Replacing the global FFT-based PSD estimate with per-frame short-time (STFT) estimates collapses OV25 EM to 20.8\% and in-distribution accuracy to 36.0\%, because the time-varying gain distorts the temporal structure of protocol waveforms that the classifier relies on.

\vspace{-0.25cm}
\section{Conclusion and Future Work}
\label{sec:conclusion}
\vspace{-0.15cm}

We presented \ourmethod{}, a framework for channel-robust wireless protocol classification under spectral overlap.
The key insight is that \textit{spectral whitening}, grounded in the scale separation between channel coherence bandwidth and OFDM subcarrier spacing, can remove channel-specific fading at the signal level without requiring any channel knowledge. Combined with a physically grounded synthetic overlap mixer that reduces dependence on real overlapping captures, \ourmethod{} achieves the single largest accuracy improvement in our evaluation on held-out OTA data while using an 889K-parameter model. Knowledge distillation further compresses the model to as few as 10K parameters for coarse spectrum awareness on 
edge devices.
Future work includes extending the framework to additional 
wireless technologies, 
and validating it across a wider range of propagation environments.




\vspace{-0.2cm}

\bibliographystyle{IEEEtran}
\bibliography{references}

\end{document}